\documentclass[a4paper,11pt]{article}
\usepackage{pos}

\usepackage{natbib}
\title{The data processing system of the POEMMA-Balloon with Radio mission  }

\author*[a,b]{Valentina Scotti}
\author[b]{Antonio Anastasio }
\author[b]{Alfonso Boiano}
\author[c]{Francesco Cafagna}
\author[b]{Vincenzo Masone } 
\author[a,b]{ Marco Mese}
\author[b]{Giuseppe Osteria}
\author[b]{Giuseppe Passeggio}
\author[a,b]{Francesco Perfetto }
\author[b]{Haroon Akhtar Qureshi }
\onbehalf{for the JEM-EUSO Collaboration\\[-1mm]{\normalsize \normalfont (a complete list of authors can be found at the end of the proceedings)}}

\affiliation[a]{Dipartimento di Fisica "E. Pancini", Università di Napoli Federico II\\
Complesso Universitario di Monte Sant'Angelo, Napoli, Italy}

\affiliation[b]{INFN, Sezione di Napoli\\
Complesso Universitario di Monte Sant'Angelo, Napoli, Italy}

\affiliation[c]{INFN, Sezione di Bari\\
Via E. Orabona 4, Bari, Italy}

\emailAdd{scottiv@na.infn.it}

\abstract{The POEMMA-Balloon with Radio (PBR) mission incorporates an advanced data processing system (DP) to enable the detection and characterization of ultra-high-energy cosmic rays and astrophysical neutrinos. The data acquisition (DAQ) system integrates inputs from the Cherenkov Camera, the Fluorescence Camera, the Radio Instrument and the X-Gamma detectors, ensuring synchronized event detection. Built upon the heritage of the EUSO-SPB2 DAQ architecture, the system has been adapted to support both the hybrid focal surface and radio instrumentation.

The DP features two redundant CPUs, differential GPS receivers, and environmental monitoring capabilities, including temperature, humidity, and gyroscope-based orientation tracking. A central clock board synchronizes data collection across all instruments, ensuring precise event reconstruction. The main trigger and clock board manages trigger signals from different detectors, supporting both joint and independent data acquisition modes. These advancements enhance the mission’s contribution to multi-messenger astrophysics and provide valuable insights for future space-based observatories.

In this paper, we describe the system’s main components and the design developed for this new mission.}

\ConferenceLogo{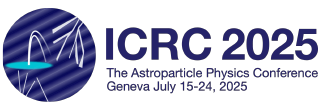}

\FullConference{39th International Cosmic Ray Conference (ICRC2025)\\
 15–24 July 2025\\
Geneva, Switzerland\\}

\begin{document}
\maketitle

\section{Introduction}

The POEMMA-Balloon with Radio (PBR) mission is a stratospheric balloon experiment designed to advance the detection of ultra-high-energy cosmic rays (UHECRs) and astrophysical neutrinos using a multi-messenger, multi-instrument approach \cite{poemma,icrc2025}. PBR combines four complementary instruments onboard a single platform: a Fluorescence Camera (FC) for observing UHECR air showers via ultraviolet fluorescence light \cite{fc}, a Cherenkov Camera (CC) optimized for detecting optical Cherenkov signals from horizontal air showers and neutrino-induced events \cite{cc}, and a Radio Instrument (RI) for capturing radio emission from extensive air showers and an X-Gamma detector to detect X-rays and gamma rays generated by the electromagnetic component of the showers \cite{pbr}. Coordinating these subsystems requires a robust, integrated data processing system capable of handling large data volumes, tight timing constraints, and the complexities of a balloon-borne observational environment.

The PBR data processor is responsible for synchronizing data acquisition across all instruments, managing trigger decisions, and ensuring efficient storage and telemetry of science data. It builds upon the architecture of the EUSO-SPB2 fluorescence DAQ system \cite{dp_spb2, spb1}, with major upgrades to accommodate the hybrid focal surface and the addition of the radio subsystem. The system includes redundant CPUs, a precision differential GPS for accurate time-stamping and localization, and a centralized clock board to distribute timing signals with sub-microsecond precision. It also features a housekeeping system that continuously monitors environmental conditions (temperature, humidity, orientation, pressure) critical for instrument calibration and performance. The data processor’s modular design enables flexible triggering logic, low-power operation, and real-time coordination of event readout, making it a central component in PBR’s effort to pioneer the detection technologies required for future space-based cosmic-ray and neutrino observatories.

\section{The DP architecture}

\begin{figure*}[bt]
\centerline{\includegraphics[width=\textwidth]{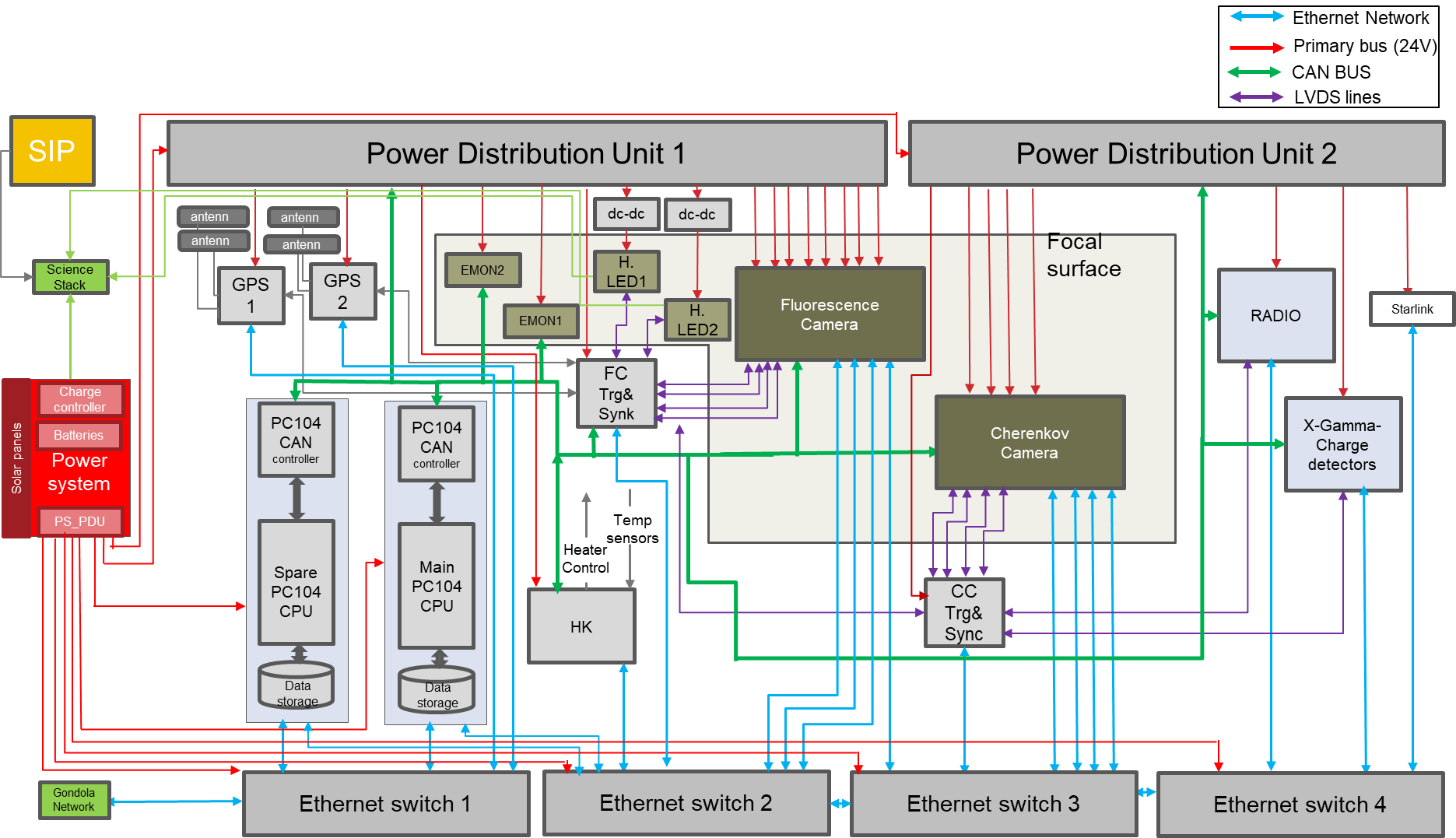}}
\caption{Block diagram of the Data Processing (DP) system for the POEMMA-Balloon with Radio (PBR) mission. The diagram illustrates the interconnection between the DP and key subsystems, including the Fluorescence Camera (FC), Cherenkov Camera (CC), Radio Instrument (RI), GPS modules, and telemetry interfaces. It highlights the modular architecture designed for redundancy, environmental monitoring, real-time data handling, and synchronization across all science instruments.}
\label{fig:blockdiag}
\end{figure*}

The block diagram of the DP and its connections with the rest of the instrument are shown in Fig.~\ref{fig:blockdiag}. 
In the following paragraphs, we will describe the Data Processing system architecture.

\subsection{Requirements}

The PBR has been designed to collect data during moonless nights to obtain the best contrast for the signatures of scientific interest. The DP links the telescope with the Gondola system. The DP is a complex system that includes most of the digital electronics of the instrument which allows a comprehensive control, configuration, monitoring, and operation of the telescope throughout the commissioning phase, test campaigns, and the entirety of the flight mission.

Besides being the interface with the flight computer, the data processing system was designed to perform the following tasks:
\begin{itemize}
        \item Main interface with Flight Computer (GCC) telemetry system
        \item Data selection/compression and transmission to Flight Computer (SIP)
        \item Power ON/OFF the whole instrument
        \item Define Telescope operation mode (Day, Night, D-N-D transitions)
        \item Configure the Front End electronics
        \item Start/Stop of the data acquisition and calibration procedures
        \item Tag events with GPS time and GPS position
        \item Synchronization of the data acquisition
        \item Manage different types of trigger signals 
        \item Manages mass memory for data storage
        \item Monitor/Control/DAQ of some Ancillary Devices
        \item Monitor voltages, current, and temperatures (LVPSs, boards, FPGAs)
\end{itemize}

The main challenges in designing and building the system are due to the physical characteristics of the signal and to the harsh environment in which the DP has to operate. Since the target flight duration for the NASA super pressure program is as long as 50 days, the requirements on electronics and data handling are quite severe. Finally, the system operated at high altitude in an unpressurized environment, which introduces a technological challenge for heat dissipation.

In addition to power and mass budget restrictions on the balloon payload, there is also a limited telemetry budget. For this reason, there is the need to prioritize data for downloading on board, besides the necessity to only record and transfer high-quality events. Everything must be done on board with minimal intervention from the ground.

\subsection{The subsystems}

The block diagram (Fig. \ref{fig:blockdiag}) shows that the data processing system is composed of several subsystems. The whole system is capable to acquire more than 8500 channels without exceeding the mass and power budget. 

Different tasks are performed by different subsystems. All the subsystems, together with the low-voltage power supply modules, are hosted in a DP box (customized Eurocard chassis) equipped with a cooling plate to dissipate heat.
\begin{itemize}
    \item CPU with Hot and Cold redundancy PCI/104 single board computer Core i7 3517UE + 5 SATA disks for data storage;
    \item HK board based on STM32 microcontroller and TIBBO EM2000; 
    \item FC Clock board \cite{clock} based on Xilinx XC7Z SoC:  manages trigger and data synchronization for the FC and also acts as an interface with the GPS receivers;
    \item CC Trigger\&Sync board based on Xilinx XC7Z SoC:  manages trigger and data synchronization for all the instruments onboard and acts as an interface with the GPS receivers;    
    \item Two GPS receivers, Trimble BX992;
    \item Ethernet switches;
    \item Solid State Power Controller.
\end{itemize}

Commercial off-the-shelf (COTS) components were predominantly selected to enhance reliability under stratospheric conditions. All the electronic components are selected to operate in an extended temperature range to enhance reliability under stratospheric conditions.

For the most critical component, the CPU, both hot and cold redundancy have been implemented. Accordingly, CAN and Ethernet protocols were selected to ensure system functionality in the event of a CPU failure.

Due to the availability of a slot in the switch in the DP, a StarLink module can also be installed. The DP guarantees the network separation requested by Columbia Scientific Balloon Facility (CSBF) and had spare PDU (Power Distribution Unit) channels available. Starlink is a low Earth orbit satellite constellation that delivers high-speed, low-latency internet. The version selected by CSBF supports download rates up to 220 Mbps.

\subsection{Time synchronization}

\begin{figure}[bt]
\centerline{\includegraphics[width=.75\linewidth]{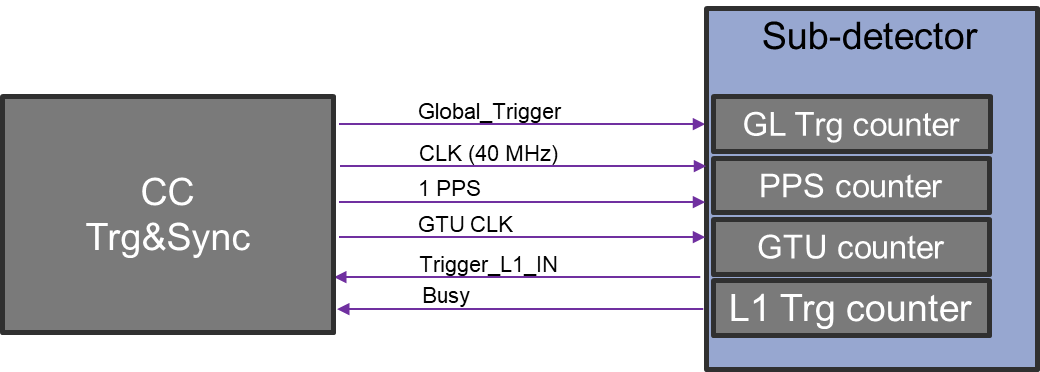}}
\caption{Schematic representation of the timing and synchronization signals within the PBR data acquisition system. The CC Trigger \& Sync board distributes the 1 PPS GPS signal, Global Trigger Unit (GTU) clock, and trigger signals to all sub-detectors. Dead time, live time, and trigger counters are included in each event packet to ensure precise inter-subsystem synchronization with sub-microsecond accuracy. This infrastructure supports coherent time-tagging across heterogeneous instruments during flight operations.}
\label{fig:time}
\end{figure}

The four PBR sub-detectors need to be synchronized by a master board (CC Trg\&Sync).
The CC Trg\&Sync board manages and distributes:
\begin{itemize}
    \item Local and global trigger signals.
    \item 1 PPS signal from GPS.
    \item GTU Clock.
    \item Busy signals from sub-detectors.
\end{itemize}
If properly managed, the time of the run should be the same for each sub-detector within one microsecond. 
Synchronization among sub-detectors is obtained through two counters: 1 PPS counter and GTU counter. The values of the two counters are stored at each Global trigger received and added to the event data, together with the values of the level 1 and Global trigger counters (see Figure \ref{fig:time}).

Each sub-detector measures its own dead and live times. For each event, the sub-detector adds to the data packet the dead time counter and the live time counter. At the end of the run, the sum of the dead and live time should be equal to the run duration.

\section{Trigger Hierarchy and Data Flow}

To manage the diverse and asynchronous signals from the three science instruments, the PBR data processor implements a hierarchical trigger architecture. At the first level, local triggers are generated independently within each subsystem: the Fluorescence Camera (FC) evaluates pixel hit patterns using SPACIROC3 ASICs; the Cherenkov Camera (CC) relies on bi-focal spatial coincidence triggers using the MIZAR ASICs and on-board FPGAs; and the Radio Instrument (RI) applies a combination of external triggers and internal radio-frequency triggers based on waveform thresholds and pattern recognition. 

Trigger requests from the CC and FC are timestamped and passed to the central data processor, which applies time-coincidence windows and acceptance logic to identify candidate events. If a trigger is validated, the processor initiates data readout from the involved instruments and assembles a unified event packet containing all relevant metadata and science data. For coincident detections across optical and radio channels, cross-instrument calibration constants and timing offsets are applied to preserve event integrity. This system enables rejection of background and anthropogenic signals while maximizing sensitivity to air showers and neutrino-induced events.

\section{Data Storage and Redundancy}

Given the limited telemetry bandwidth available on a balloon platform, PBR employs onboard solid-state storage to archive full-resolution science data. The data processor includes two redundant central processing units, each capable of buffering event data, managing file systems, and handling prioritized data queues for downlink. Event data are stored locally with robust time-stamping and metadata annotation to facilitate offline analysis. The system includes failure detection and automatic switchover logic to maintain data integrity in the event of component failure.

Storage prioritization is implemented to ensure that high-value events, such as upward-going EAS candidates or Target of Opportunity-related observations \cite{too}, are retained even during high event rates. Lower priority data may be downsampled, compressed, or flagged for later deletion if storage capacity becomes constrained during long-duration flights.

\section{Telemetry and Communications}

The PBR payload is equipped with a multi-tiered telemetry system. Low-bandwidth communication for command and control is maintained via an Iridium satellite link (255 bytes/minute), while science data is downlinked primarily through NASA's Tracking and Data Relay Satellite System (TDRSS), offering up to 130~kbps under optimal conditions. In addition, the payload is equipped with two high-rate Starlink maritime terminals for redundant bulk data transmission, which have been tested on recent NASA balloon flights \cite{spb2}.

During flight, the data processor coordinates data packaging and scheduling for transmission through these channels based on available bandwidth and pre-programmed prioritization. Critical housekeeping and quick-look science summaries are continuously downlinked for ground-based health monitoring and near-real-time assessment of mission performance. Slow control (on the CAN bus) is used to control and acquire HK data from sub-systems PDUs, EMONs, and HK board. The complete dataset is preserved onboard and retrieved upon payload recovery, or streamed in segments via high-speed links as telemetry allows.

\section{The software}

The DP functions as an interface, forging a critical link between the detector and the telemetry blocks, while also facilitating seamless interaction with the end users. Instrument operations and monitoring will be shared between collaborators in the US, Japan, and Europe to provide, during local daytime, control and monitoring of the telescopes 24/7, similarly to the on-ground control that was done for EUSO-SPB2 \cite{software}.  

The CPU software can be divided into two categories:
\begin{itemize}
    \item the control software 
    \item the data handling software
\end{itemize}

The control software exhibits a high level of flexibility, crafted to adapt to various phases encompassing both commissioning and the flight mission. Notably, this software enables control and monitoring of the detector, while also ensuring comprehensive data storage, encompassing scientific data, housekeeping records, and log information. The design enables full user control via telemetry, enabling real-time command and supervision. Significantly, during the flight mission, it is imperative to maintain continuous surveillance of the instrument's status. Such vigilant monitoring is essential to discern the opportune moments for initiating or halting measurements. The software's adeptness in facilitating constant status checks proved paramount in making decisive determinations, enhancing the precision and efficiency of the overall mission operations.

The data handling software manages science data acquisition from the CLK and Zynq boards. The acquisition software initializes and configures the subsystems, monitors the connections with the CPU, and verifies the behavior of the subsystems. It can perform several types of acquisitions, such as recording externally triggered events or internally triggered events. Finally, it manages the prioritization folder structure for down-link to cope with the maximum amount of telemetry available each day.

Science Stack data will be available in real time and will be used to monitor and control the status of the telescope. In particular, before starting the power-on or power-off sequence, it was necessary to monitor:
\begin{itemize}
    \item Status of the power system;
    \item Temperatures of critical points of the telescope (Batteries, mirror, focal surface, electronics, etc.);
    \item Light level of the EMON (Day/Night transition, Night/Day transition);
    \item Status of the aperture shutter (Open/Close).
\end{itemize}

\section{Conclusion}

The DP system developed for the POEMMA-Balloon with Radio mission represents a significant advancement in the design and integration of multi-instrument data acquisition platforms for high-altitude astrophysical observations. Building upon the heritage of EUSO-SPB2, the PBR DP introduces key innovations to support a hybrid focal surface that includes fluorescence, Cherenkov, and radio detection capabilities.

The architecture features both hot and cold CPU redundancy, a flexible and hierarchical triggering system, sub-microsecond synchronization across all subsystems, and robust solid-state data storage tailored for long-duration balloon flights. The adoption of reliable communication protocols (CAN and Ethernet) and environmental monitoring systems ensures autonomous, resilient operation under the challenging conditions of a stratospheric environment.

The modular design and real-time telemetry management enable efficient science data prioritization, onboard event selection, and robust system health monitoring. These characteristics make the PBR DP system not only a critical component of the current balloon-borne mission, but also a testbed for future space-based cosmic ray and neutrino observatories.

Continued testing and integration will refine its performance ahead of flight operations, with anticipated contributions to both scientific output and the development of next-generation multi-messenger platforms.

\section{Acknowledgements}

The authors acknowledge the support by NASA award 80NSSC22K1488 and 80NSSC24K1780, by the French space agency CNES and the Italian space agency ASI. The work is supported by OP JAC financed by ESIF and the MEYS $CZ.02.01.01/00/22\_008/0004596$. This research used resources of the National Energy Research Scientific Computing Center (NERSC), a U.S. Department of Energy Office of Science User Facility operated under Contract No. DE-AC02-05CH11231. We also acknowledge the invaluable contributions of the administrative and technical staff at our home institutions.

    \newpage
{\Large\bf Full Authors list: The JEM-EUSO Collaboration}

\begin{sloppypar}
{\small \noindent
M.~Abdullahi$^{ep,er}$              
M.~Abrate$^{ek,el}$,                
J.H.~Adams Jr.$^{ld}$,              
D.~Allard$^{cb}$,                   
P.~Alldredge$^{ld}$,                
R.~Aloisio$^{ep,er}$,               
R.~Ammendola$^{ei}$,                
A.~Anastasio$^{ef}$,                
L.~Anchordoqui$^{le}$,              
V.~Andreoli$^{ek,el}$
% ...
}
\end{sloppypar}
\vspace*{.3cm}

{ \footnotesize
\noindent
%
% ...
$^{cb}$ Universit\'e de Paris, AstroParticule et Cosmologie (CNRS), Paris, France\\
% ...
$^{ef}$ Istituto Nazionale di Fisica Nucleare (INFN), Sezione di Napoli, Naples, Italy\\
% ...
$^{ei}$ Istituto Nazionale di Fisica Nucleare (INFN), Sezione di Roma Tor Vergata, Rome, Italy\\
% ...
$^{ek}$ Istituto Nazionale di Fisica Nucleare (INFN), Sezione di Torino, Turin, Italy\\
$^{el}$ Universit\`a di Torino, Dipartimento di Fisica, Turin, Italy\\
% ...
$^{ep}$ Gran Sasso Science Institute (GSSI), L’Aquila, Italy\\
$^{er}$ Istituto Nazionale di Fisica Nucleare (INFN), Laboratori Nazionali del Gran Sasso, Assergi, Italy\\
% ...
$^{ld}$ University of Alabama in Huntsville, Huntsville, AL, USA\\
$^{le}$ City University of New York (CUNY), Lehman College, Bronx, NY, USA\\
% ...
}

\end{document}